\documentclass[%
 reprint,
superscriptaddress,
 amsmath,amssymb,
prl,
]{revtex4-1}

\usepackage{graphicx}
\usepackage{dcolumn}
\usepackage{bm}
\usepackage{color}
\usepackage[english]{babel}

\begin{document}


\title{Magnon diffuse scattering in scanning transmission electron microscopy}

\author{Keenan Lyon}
\email{keenan.lyon@physics.uu.se}
\affiliation{Department of Physics and Astronomy, Uppsala University, L\"{a}gerhyddsv\"{a}gen 1, Uppsala, Sweden}

\author{Anders Bergman}%
\affiliation{Department of Physics and Astronomy, Uppsala University, L\"{a}gerhyddsv\"{a}gen 1, Uppsala, Sweden}

\author{Paul Zeiger}
\affiliation{Department of Physics and Astronomy, Uppsala University, L\"{a}gerhyddsv\"{a}gen 1, Uppsala, Sweden}

\author{Demie Kepaptsoglou}%
\affiliation{SuperSTEM Laboratory, SciTech Daresbury Campus, Daresbury WA4 4AD, United Kingdom}
\affiliation{Department of Physics, University of York, York YO10 5DD, United Kingdom}

\author{Quentin M. Ramasse}%
\affiliation{SuperSTEM Laboratory, SciTech Daresbury Campus, Daresbury WA4 4AD, United Kingdom}
\affiliation{School of Chemical and Process Engineering, University of Leeds, Leeds LS2 9JT, United Kingdom}%
\affiliation{School of Physics and Astronomy, University of Leeds, Leeds LS2 9JT, United Kingdom}

\author{Juan Carlos Idrobo}
\affiliation{Center for Nanophase Materials Sciences, Oak Ridge National Laboratory, Oak Ridge, TN, USA}

\author{J\'{a}n Rusz}
\email{jan.rusz@physics.uu.se}
\affiliation{Department of Physics and Astronomy, Uppsala University, L\"{a}gerhyddsv\"{a}gen 1, Uppsala, Sweden}


\date{\today}

\begin{abstract}
We present a theory and a simulation of thermal diffuse scattering due to the excitation of magnons in scanning transmission electron microscopy. The calculations indicate that magnons can present atomic contrast when detected by electron energy-loss spectroscopy using atomic-size electron beams. The  results presented here indicate that the intensity of the magnon diffuse scattering in bcc iron at 300~K is 4 orders of magnitude weaker than the intensity of thermal diffuse scattering arising from atomic vibrations. In an energy range where the phonon and magnon dispersions do not overlap, a monochromated scanning transmission electron microscope equipped with direct electron detectors for spectroscopy is expected to resolve the magnon spectral signatures.
\end{abstract}

\maketitle

Efficient electron beam monochromators have extended the already wide range of experimental techniques available to scanning transmission electron microscopy (STEM) by allowing for the measurement of vibrational spectra \cite{krivanek_vibrational_2014,krivanek_progress_2019}. Since its first demonstration, vibrational electron energy-loss spectroscopy (EELS) is developing at a swift pace. Several key milestones have been reached, such as the identification of isotope compositions \cite{hachtel_identification_2019}, the detection of atomic level contrast in vibrational signals \cite{hage_phonon_2019,venkatraman_vibrational_2019} and of the spectral signatures of an individual impurity atom \cite{hage_single-atom_2020}, and spatial- and angle-resolved measurements on a single stacking fault \cite{yan_stacking_fault_2021}. Such vibrational modes occupy an energy range from zero to tens or few hundreds of meV in solid state materials. Qualitatively the same energy range is also occupied by energy losses due to excitations of magnons, arising from the collective excitation of the electrons' spin in a lattice.

Magnons represent collective excitations of the magnetic subsystem and semi-classically they can be visualized as a wave of precessing magnetic moments~\cite{atomistic_spin_dynamics_2017}. Among other inelastic scattering experimental techniques~\cite{Sinha1969,Keatley2017,Sebastian2015}, magnons can be efficiently excited by electron scattering in reflection geometry: spin-polarised EELS (SPEELS) or reflection (REELS) using spin- and non-polarised electron sources, respectively  \cite{VOLLMER20042126,ZAKERI2013157,Ibach2017}. It is therefore expected that, in direct analogy with phonon spectroscopy, the spectroscopic signature of
magnons and their dispersion in momentum space should also be accessible within the remit of vibrational STEM-EELS. Due to the scattering cross sections and low penetration depth for low energy electrons (typically the incident beam energy does not exceed 10~eV), SPEELS (REELS) measurements are restricted to the detection of spin waves at surfaces or ultra‐thin films. However, although these techniques allow the probing of magnon excitations and their energy‐momentum dispersion with high‐energy resolution, localised information stemming, for example, from buried interfaces or point defects, is lost. The promise, therefore, of probing magnons with STEM-EELS at simultaneous high spatial and energy resolution is highly attractive. 

It is well-known that the interaction of magnetic moments with the electron beam is significantly weaker than its interaction with the Coulomb potential, often by 3 or 4 orders of magnitude \cite{rother_relativistic_2009,loudon_antiferromagnetic_2012}. Since magnons represent time-dependent distortions of the magnetic structure, similarly as phonons represent time-dependent distortions of the crystal structure, one might conclude that the inelastic magnon signal will be of 3 to 4 orders of magnitude weaker than the phonon signal, which would certainly make their detection very challenging. In certain contexts, however, the magnetic effects in the elastic scattering regime can reach relative strengths of up to a few percent \cite{edstrom_vortex_prl_2016,edstrom_vortex_prb_2016,edstrom_dpc_2019}.

Furthermore, the use of direct or hybrid-pixel detectors for spectroscopy applications already offers drastically improved detection dynamic range and low background noise, with signals a mere $10^{-7}$ of the full beam intensity readily detectable within tens of pixels of the recorded signal's maximum \cite{plotkin-swing_hybrid_2020}. Together with improvements in monochromator and spectrometer design, resulting in increased energy resolutions in particular at lower acceleration voltages (4.2~meV at 30~kV \cite{krivanek_progress_2019}), where the inelastic cross-sections are more favorable, signals of weak intensity such as energy losses due to excitation of magnons could be accessible experimentally.

In this Letter, we address the prospects for detection of magnons in monochromated STEM. For this purpose we devise a model for the simulation of thermal diffuse scattering (TDS) due to magnons. Consequent simulations of the inelastic magnon scattering show an intensity distribution of the magnon signal in the diffraction plane under studied experimental conditions. We compare the intensity of magnon scattering with that of phonon scattering. We demonstrate that atomic resolution spectrum imaging with a magnon scattering signal is possible. On the basis of this information, we discuss the feasibility of the experimental conditions for the detection of an atomically resolved magnon signal, in the context of currently available instrumentation and of further technical developments known to be achievable in the near future.

First, we introduce our theoretical model for the calculation of the magnon TDS. The model is based on an analogy with the so-called quantum excitation of phonon (QEP) model \cite{forbes_quantum_2010}, where it was argued that the inelastic signal due to atomic vibrations can be accessed via sampling from all possible atomic displacement configurations. Averaging over intensities (squared amplitudes) of exit wave-functions results in an incoherent (total) scattering intensity. Instead, taking the squared amplitude of the averaged exit wavefunction results in a coherent (elastic) scattering intensity. The magnon inelastic scattering signal is then the difference between the incoherent and coherent scattering intensities. Instead of vibrating atoms, we deal here with wiggling magnetic moments. They cause local deviations of the microscopic magnetic fields from their time-average. These deviations influence the electron beam propagating through the sample and it is what allows the detection of magnons with the electron microscope. 

To realize this formal analogy practically, we need three key components. The first component is a beam propagation method that can take into account the influence of a microscopic magnetic field on the electron beam. Such a method exists and has been described in Refs.~\cite{rother_relativistic_2009,edstrom_vortex_prl_2016,edstrom_vortex_prb_2016}. In works of Edstr\"{o}m et al., a Pauli multislice method was introduced, which is a generalization of the classical multislice method \cite{cowley_scattering_1957}, deriving a paraxial approximation starting from Pauli's equation rather than Schr\"{o}dinger's equation.

The second component is a method for generating the microscopic magnetic field $\mathbf{B(r)}$ and microscopic magnetic vector potential $\mathbf{A(r)}$. In Refs.~\cite{rother_relativistic_2009,edstrom_vortex_prl_2016,edstrom_vortex_prb_2016,edstrom_dpc_2019}, $\mathbf{A(r)}$ and $\mathbf{B(r)}$ were generated from spin-densities evaluated by calculations within the density functional theory framework. However, for large supercells containing many thousands of atoms with varying orientations of magnetic moments, such an approach becomes impractical. For this purpose we have developed a parametrization of the magnetic fields and vector potentials \cite{keenan_parametrization_2021}. 

The parametrization starts  with the microscopic magnetic vector potentials and fields of single atomic systems, where a quasi-dipole model is fitted for each individual element in a way analogous to electron form factors while also accounting for the reduced magnetic moment in the atomic to bulk transition. 
As the contributions to these magnetic quantities from each atom can simply be summed up in superposition, knowledge of the positions and moment directions for a supercell in tandem with this parametrization allows for efficient computation of these magnetic quantities in systems of any size.

The third and final component provides realistic snapshots of precessing magnetic moments in the supercell. This can be efficiently realized by semi-classical atomistic spin dynamics (ASD) simulations \cite{atomistic_spin_dynamics_2017}. ASD is a magnetic analogue of how molecular dynamics (MD) simulations describe atomic vibrations, here providing a realistic description of thermally excited magnetic configurations and their dynamics. Here we have used the \texttt{UppASD} code \cite{uppasd_url} for the ASD simulations. 

The parameter settings for each step are described in reverse order. As our model system we chose ferromagnetic bcc iron, a prototypical system for \textit{ab initio} calculations of magnetic properties, and one where magnons have been previously detected experimentally by energy loss spectoscopy using SPEELS \cite{ZAKERI2013157}. A supercell of $30 \times 30 \times 100$ unit cells containing 180000 atoms, with dimensions $86 \times 86 \times 287$~\AA{}$^3$, has been constructed. The ASD method relies on a parameterized spin Hamiltonian to describe the spin dynamics. Here we have used a Heisenberg Hamiltonian defined by interatomic exchange interactions calculated \textit{ab initio} using the SPRKKR\cite{sprkkr_2011} code. With this realistic description of the spin dynamics in bcc Fe, we performed ASD simulations in order to get representative snapshots of the magnetization in the sample with a sampling interval of 1 ps. The simulations were performed with a timestep of 0.1 fs and we used a large Gilbert damping parameter of 0.1 in order to minimize the correlation between different snapshots. 

In the generation of parametrized fields we utilize a numerical grid of $1500 \times 1500 \times 3000$ grid points spanning the entire supercell, with points within a $6 \times 6 \times 6$~\AA{}$^3$ cube surrounding each atom evaluated for the magnetic vector potentials and magnetic fields. 

The Pauli's-equation-based multislice calculations have been done on the same numerical grid.
The cut-off for atomic potentials, using Kirkland's parametrization \cite{kirkland_advanced_2010}, was set to 4~\AA{}. Zero aberrations for the incoming wave have been assumed, and the beam was focused on the entrance surface of the supercell.

\begin{figure}
    \centering
    \includegraphics[width=\columnwidth]{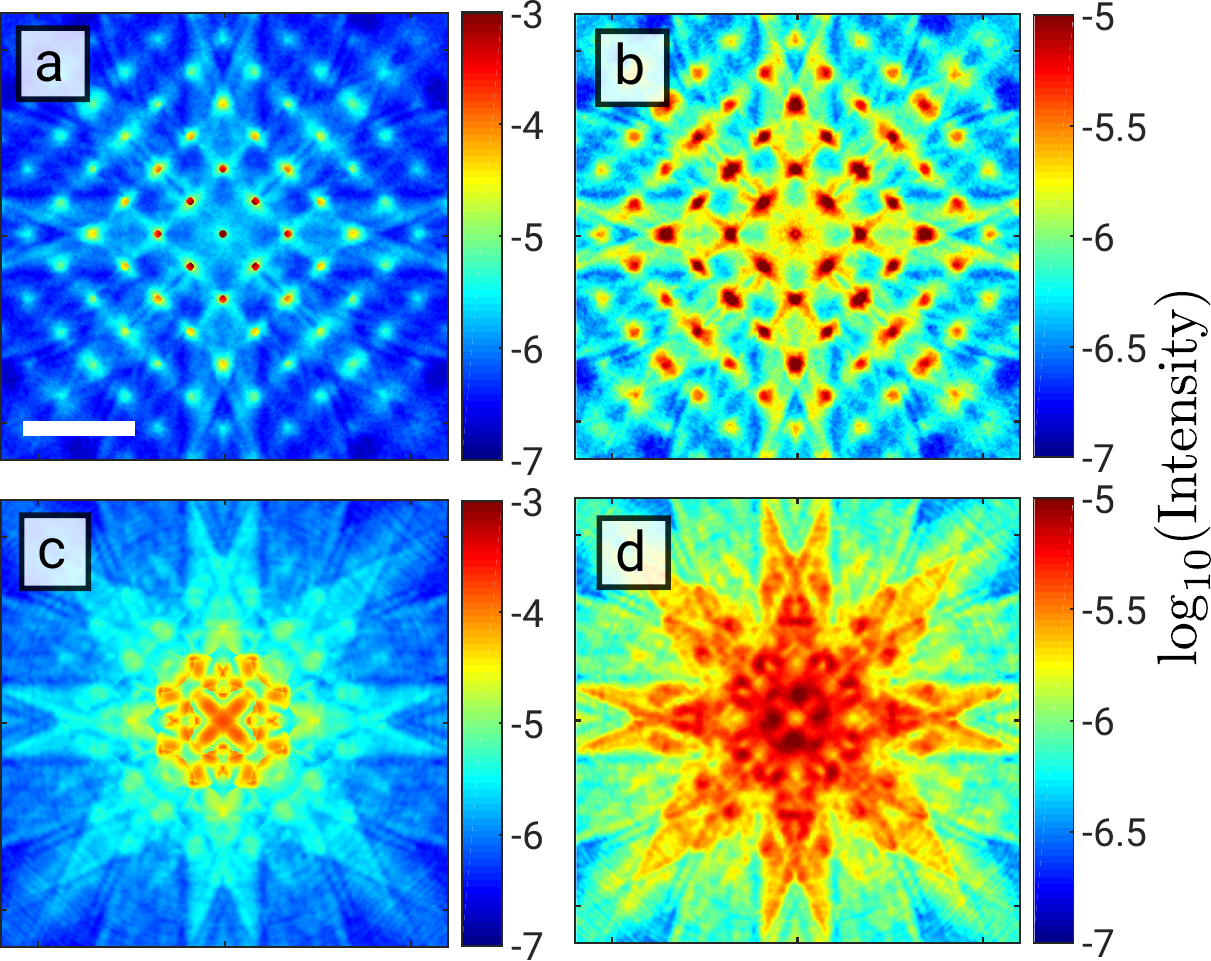}
    \caption{Thermal diffuse scattering simulation due to atomic vibrations. Acceleration voltage was set to 200~kV. This calculation neglects magnetic effects. Convergence semi-angle was set to 1~mrad in a) and b), and to 10~mrad in c) and d). The total diffraction pattern is shown in panels a) and c), while the pure phonon signal is shown in panels b) and d). Intensity is plotted on a logarithmic scale for scattering angles of $\pm 60$~mrad along both axes, with white bar in panel a) representing 30~mrad.}
    \label{fig:phononTDS}
\end{figure}

Since the magnon TDS will create nonzero scattered intensity in between Bragg spots, it will overlap with the TDS stemming from atomic vibrations. It is therefore important to know the distribution and intensity of the phonon TDS signal in the diffraction plane. We compute the phonon TDS in a similar fashion to the magnon TDS, but instead of snapshots of the time-varying magnetic field, snapshots of the vibrating crystal structure are required. To that end, we simulate the atomic vibrations with the \texttt{LAMMPS} MD code \cite{plimpton_fast_1995,lammpsweb}. An orthogonal simulation box consisting of $30 \times 30 \times 100$ unit cells of bcc-Fe was considered with a lattice parameter $a=2.859$~\AA{}, which was determined from the time average of the simulation box dimensions via isothermal-isobaric MD simulation at a temperature of 300~K and a pressure of 0.0~bar. In all MD simulations, the time step was set to $0.001$~ps and the embedded-atom method potential described the inter-atomic interactions \cite{mendelev_development_2003}. One MD trajectory was simulated using a Langevin thermostat at a temperature of 300~K, from which one structure snapshot was taken every 2000 time steps, i.e., every 2~ps, for a total of 96 snapshots. Figure~\ref{fig:phononTDS} shows a calculated total (incoherent) diffraction pattern and its inelastic phonon component. The relative strength of energy-integrated phonon signal intensity reaches up to $10^{-2}$ of the total intensity. 

\begin{figure}
    \centering
    \includegraphics[width=\columnwidth]{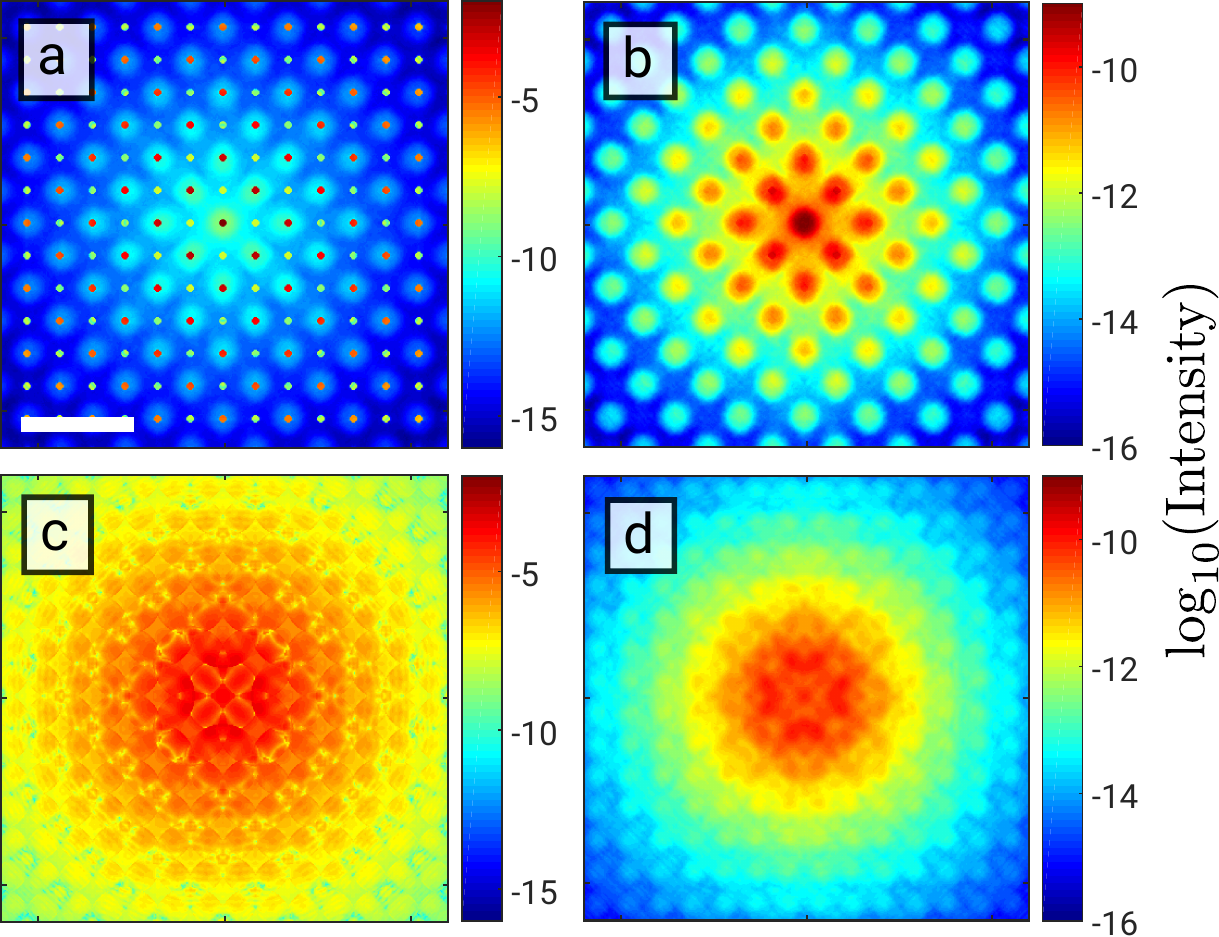}
    \caption{Diffuse scattering simulation due to precession of magnetic moments, i.e., excitation of magnons. Atomic vibrations are neglected in this simulation. Convergence semi-angle was set to 1~mrad in a) and b), and to 10~mrad in c) and d). The total diffraction pattern is shown in panels a) and c), while pure magnon signal is shown in panels b) and d). Intensity is plotted on a logarithmic scale for scattering angles of $\pm 60$~mrad along both axes, with white bar in panel a) representing 30~mrad.}
    \label{fig:DP1mrad}
\end{figure}

Recent high-energy-resolution STEM-EELS experiments in the vibrational regime have explored the use of either nano-probe or atomic resolution modes. Nano-probe offers nanometer-scale spatial resolution alongside interpretable angle-resolved measurements at moderate momentum resolution \cite{hage_phonon_2019}. In such a case, small convergence semi-angles are being used, such as 1~mrad. In atomic resolution mode we form an \AA{}ngstr\"{o}m-sized electron probe by using substantially larger convergence semi-angles, such as 20--30~mrad. Experiments with atomic size electron beams offer atomic-resolution EELS but without momentum sensitivity \cite{yan_stacking_fault_2021,hage_single-atom_2020,venkatraman_vibrational_2019}. Given the practical and conceptual similarities between these vibrational spectroscopy results and the experiments envisaged here to study magnons, these two use-cases will form the basis for the choice of simulation parameters.

We start our discussion of magnon TDS with simulations assuming a nanometer-sized electron beam with a convergence semi-angle of 1~mrad at an acceleration voltage of 200~kV, see Fig.~\ref{fig:DP1mrad}. Although lower acceleration voltages, typically 30-60~kV, have been used in recent vibrational and ultra-low loss STEM-EELS experiments, due in particular to higher inelastic scattering cross-sections, the sample requirements (e.g. the risk of oxidation in very thin lamellae of bbc-Fe) make the choice of 200~kV pertinent as it would allow for the observation of thicker objects. Panel a) shows the central part of the diffraction pattern within a $\pm 60$~mrad range of scattering angles, where the atomic vibrations have been neglected. It can be seen that the intensity of the diffuse signal due to the excitation of magnons is of a similar order of magnitude as forbidden reflections, that is approximately $10^{-6}$--$10^{-7}$ of the transmitted beam intensity. In a simulation including atomic vibrations the forbidden reflections are not visible, being dominated by the vibrational TDS, see Fig.~\ref{fig:phononTDS}a). The pure inelastic signal component due to magnons is shown in panel b). It forms clouds of intensity centered around the Bragg reflections. The per-pixel maximum relative intensity of the magnon and phonon TDS reaches $3 \times 10^{-4}$. Panels c) and d) show an analogous calculation for an electron beam with convergence semi-angle of 10~mrad. The relative intensity of the magnon TDS remains at a very similar level. Figures~\ref{fig:DP1mrad}b) and d) represent the energy-integrated magnon EELS signal, i.e., excluding the zero-loss peak and phonon EELS intensity. We have performed similar calculations for a 5~nm thick sample at 30~kV acceleration voltage (not shown) with qualitatively similar outcomes.

\begin{figure}
    \centering
    \includegraphics[width=\columnwidth]{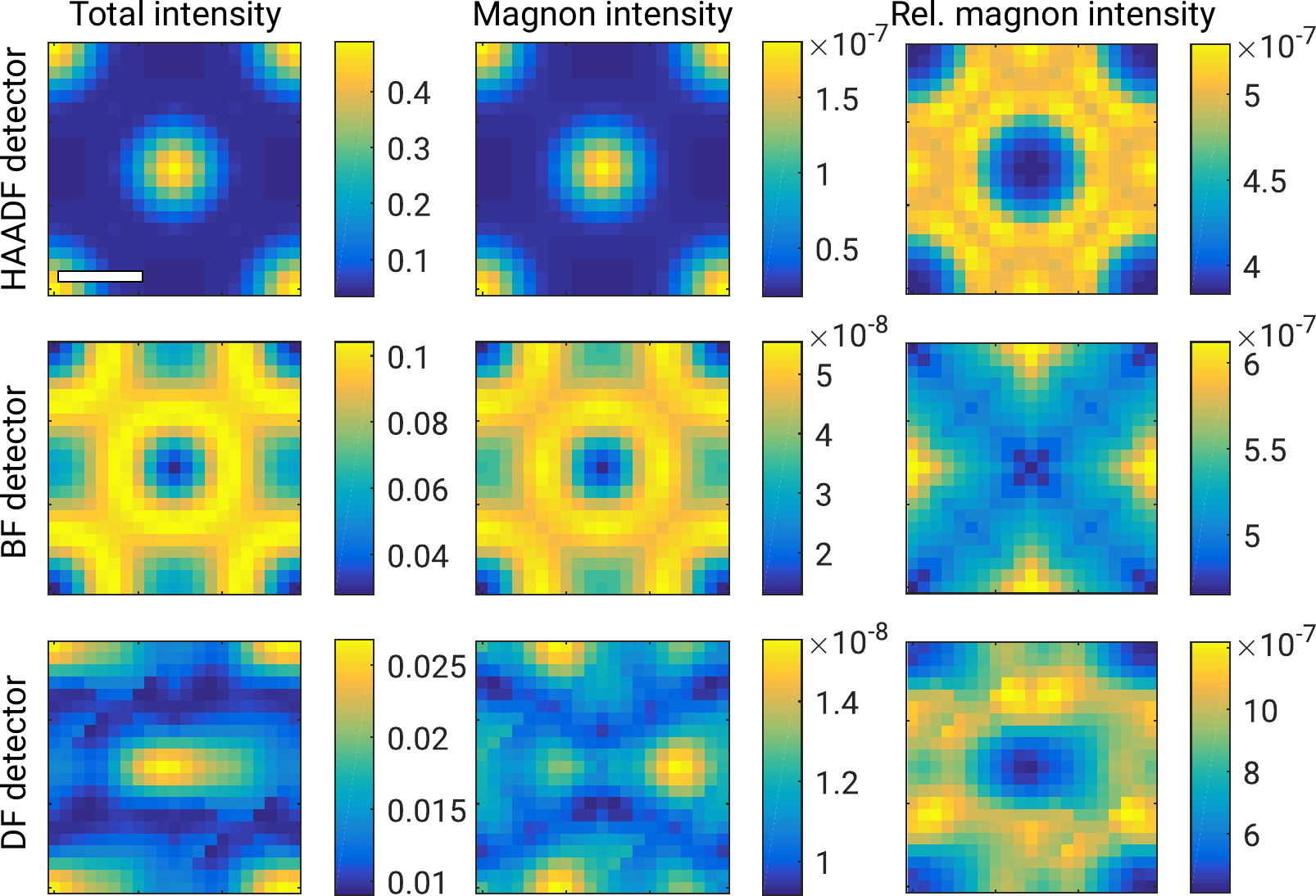}
    \caption{STEM images calculated with an electron beam accelerated by 100~kV and convergence semi-angle of 30~mrad, while neglecting the atomic vibrations. White bar in the top left panel marks 1~\AA{}. Individual rows correspond to three different detector setups: HAADF detector with inner/outer collection semi-angles of 100~mrad and 250~mrad, BF detector with collection semi-angle of 30~mrad and a DF detector displaced by 60~mrad off-axis along the $\theta_x$ direction, with a collection semi-angle of 30~mrad. Individual columns correspond to the total intensity (excluding phonon contribution), magnon scattering intensity, and their ratio.}
    \label{fig:magnonSTEM}
\end{figure}

For convergence semi-angles of approximately 10~mrad and beyond, the Bragg discs already overlap, meaning that we are in the atomic resolution regime. An intriguing question arises, whether atomic scale contrast could be reached using purely the magnon EELS signal. In Fig.~\ref{fig:magnonSTEM} we present a calculation of a STEM image based purely on the inelastic magnon intensity. A convergence semi-angle of 30~mrad and acceleration voltage of 100~kV have been assumed. We have considered three typical detector settings: high-angle annular dark field (HAADF) with inner/outer collection semi-angles of 100~mrad and 250~mrad, bright field (BF) with collection semi-angle of 10~mrad and  off-axis dark field (DF) detector with collection semi-angle of 30~mrad, displaced by 60~mrad from the center of diffraction pattern along $\theta_x$-axis. These could be thought of as energy-filtered STEM images, that is, spectrum images collected using equivalent collection angle ranges, but generated by integrating the energy range solely over magnon losses. An experimental realization may thus be possible if the most intense magnon peaks are sufficiently separated from other losses in the corresponding energy-loss range. 

For all three detector settings we observe an atomic scale contrast, both in total scattering intensity (left column) as well as in the magnon diffuse scattering (middle column). The HAADF detector leads to the highest atomic level contrast, both in total and in magnon scattering intensity. On the other hand, owing to the very similar spatial distribution of both signals, the ratio of the magnon signal intensity to the total scattering intensity (right column) remains under $10^{-6}$. The magnon signal of the BF detector shows expected strong dynamical diffraction effects with volcano-shaped features around the atomic columns. A similar behavior was reported in phonon EELS maps \cite{HagePRB2020}. The contrast is lower with a ratio of maximal to  minimal intensities near a factor of 5, while the relative strength of the magnon signal remains below $10^{-6}$, at a similar level to the magnon signal collected by the HAADF detector. For the chosen geometry of the off-axis DF detector, in the total scattering cross-section we observe elongated features at the position of atomic columns. Interestingly, the peaks of the magnon signal are found to be strongly displaced to the right, by about 0.8~\AA{} in the direction parallel to the displacement of the detector. Similar observations were made about phonon EELS in Ref.~\cite{hBN_APB_2021}, albeit to a smaller extent. Thanks to the qualitatively different STEM image of the total vs magnon signal, the relative strength of magnon scattering intensity is higher here, reaching $1.3 \times 10^{-6}$.

Although this level of signal is small in absolute terms, recent technological progress makes their detection a very realistic prospect. For instance, the vibrational signature of guanine molecules obtained using a hybrid-pixel detector revealed peaks of intensity well below $2 \times 10^{-5}$ that of the zero-loss peak (ZLP), across an energy-loss range starting as low as $\sim20$~meV. \cite{dellby_HREELS_Abstract} Thanks to the increased energy resolution of new-generation monochromators, the numerous peaks in the low energy-loss sector are also well-separated, making it easier to isolate the contribution from each allowed vibrational mode into, e.g., an inelastic `phonon' image or `magnon' image, as simulated here. Furthermore, in a dark-field geometry, such as that described above for the HAADF or off-axis DF detectors, the relative intensity of the ZLP is vastly reduced (for a lower overall signal) and its intensity for dark-field phonon imaging is of a similar order of magnitude to that of vibrational peaks. \cite{hage_phonon_2019} This alternative electron optical geometry should allow for an enhancement of the relative magnon signal and its detection in future experiments, especially if direct electron detectors are used.

In the presented simulations it was assumed that the sample is subjected to a 1~T magnetic field parallel to the beam coming from the objective lens. Newly-developed instrumentation makes it possible to either null or control the magnetic field of the STEM objective lens at the sample while retaining an atomic-size probe. \cite{Shibata-MARS} It would thus be of interest to analyze the impact of the strength and direction of the magnetic field on the observed magnon signal as an additional experimental means to boost or isolate magnon peaks in the recorded spectra. Another parameter of great interest is the temperature. It is likely that the temperature dependence of the phonon and magnon signals will qualitatively differ, perhaps even offering the possibility for separation of these signals. Bcc iron has a magnetic transition temperature above 1000~K, therefore one would expect stronger magnon scattering intensities at higher than room temperatures. All these aspects will be the subject of future research.


The simulations presented here lack the spectroscopic dimension in the magnon scattering: all calculations are integrated over all possible energy losses where magnons contribute to the spectra. We aim to extend our work to include energy channel sensitivity by analogy with the frequency-resolved frozen phonon multislice method (FRFPMS; \cite{zeiger_efficient_2020,zeiger_plane_wave_2021}). For that purpose, we are implementing colored thermostats into \texttt{UppASD}. Such simulations will provide snapshots of magnetic moments vibrating only within desired frequency ranges. This will allow us to assemble magnon EEL spectra by repeating the simulation procedure described here for each of the frequency ranges.

In summary, we have constructed a model and presented simulations of magnon thermal diffuse scattering. The magnon scattering has a relative intensity of up to $10^{-4}$ of the phonon scattering intensity, suggesting that for an initial detection of magnons in experiments one should use samples with a magnon energy position and dispersion that is sufficiently separated from that of the phonons. On top of this, an optimal sample would be a magnetic insulator or semiconductor with a bandgap wider than the width of the magnon spectrum. Under these conditions, we consider the detection of magnons in STEM-EELS experiments is a realistic goal for suitable material systems. The successful fingerprinting of magnons in an electron microscope will create a radical new way of studying the fundamentals of magnetic ordering and spin wave excitations, e.g. in  material systems used for spintronics and spin caloritronics, where spin currents are propagated by magnons \cite{Sinova2015}.

\begin{acknowledgments}
We acknowledge the support of Swedish Research Council and Carl Trygger's Foundation. Simulations were enabled by resources provided by the Swedish National Infrastructure for Computing (SNIC) at NSC Centre partially funded by the Swedish Research Council through grant agreement no.\ 2018-05973. SuperSTEM is the National Research Facility for Advanced Electron Microscopy funded by the Engineering and Physical Sciences Research Council (EPSRC). We acknowledge financial support from the Engineering and Physical Sciences Research Council (EPSRC) via grant no EP/V048767/1. JCI acknowledges support of the Center for Nanophase Materials Sciences, which is a DOE Office of Science User Facility. 
\end{acknowledgments}

\bibliographystyle{apsrev4-1}
\bibliography{references}

\end{document}